\documentclass[prd,twocolumn,nofootinbib,showpacs]{revtex4-1}

\usepackage{amsfonts}
\usepackage{amsmath}
\usepackage{amssymb}
\usepackage{bm}
\usepackage{dcolumn}
\usepackage{graphicx}
\usepackage{graphics}
\usepackage[latin1]{inputenc}
\usepackage{latexsym}
\usepackage{rotating}
\usepackage{url}
\usepackage{xspace} 
\usepackage[usenames]{color}
\usepackage{mathrsfs}
\usepackage{hyperref}

\definecolor {darkgreen}{rgb}{0.2,0.7,0.2}

\newcommand\be{\begin{equation}}
\newcommand\ba{\begin{eqnarray}}
\newcommand\ee{\end{equation}}
\newcommand\ea{\end{eqnarray}}

\newcommand\bw{\begin{widetext}}
\newcommand\ew{\end{widetext}}

\newcommand{\nn}{\nonumber}

\newcommand{\Schw}{{\mbox{\tiny Schw}}}
\newcommand{\Kerra}{{\mbox{\tiny Kerr,$\chi^1$}}}
\newcommand{\Kerraa}{{\mbox{\tiny Kerr,$\chi^2$}}}
\newcommand{\QGpert}{{\mbox{\tiny QG}}}
\newcommand{\QG}{{\mbox{\tiny QG,$\chi^0$}}}
\newcommand{\QGa}{{\mbox{\tiny QG,$\chi^1$}}}
\newcommand{\QGaa}{{\mbox{\tiny QG,$\chi^2$}}}
\newcommand{\Deltaaa}{{\mbox{\tiny $\Delta,\chi^2$}}}
\newcommand{\mat}{{\mbox{\tiny mat}}}

\newcommand{\GR}{{\mbox{\tiny GR}}}

\newcommand{\CS}{{\mbox{\tiny CS}}}

\newcommand{\EDGB}{{\mbox{\tiny EDGB}}}

\begin{document}
\title{Linear Stability Analysis of Dynamical Quadratic Gravity}

\author{Dimitry Ayzenberg}
\affiliation{Department of Physics, Montana State University, Bozeman, MT 59717, USA.}

\author{Kent Yagi}
\affiliation{Department of Physics, Montana State University, Bozeman, MT 59717, USA.}

\author{Nicol\'as Yunes}
\affiliation{Department of Physics, Montana State University, Bozeman, MT 59717, USA.}

\date{\today}

\begin{abstract} 
We study the linear stability of dynamical, quadratic gravity, focusing on two particular subclasses (the even-parity sector, exemplified by Einstein-Dilaton-Gauss-Bonnet gravity, and the-odd parity sector, exemplified by dynamical Chern-Simons modified gravity) in the high-frequency, geometric optics approximation. This analysis is carried out by studying gravitational and scalar modes propagating on spherically symmetric and axially symmetric, vacuum solutions of the theory and finding the associated dispersion relations. These relations are solved in two separate cases (the scalar regime and the gravitational wave regime, defined by requiring the ratio of the amplitude of the perturbations to be much greater or smaller than unity) and found in both cases to not lead to exponential growth of the propagating modes, suggesting linearly stability. The modes are found to propagate at subluminal and superluminal speeds, depending on the propagating modes' direction relative to the background geometry, just as in dynamical Chern-Simons gravity.
\end{abstract}

\maketitle

\section{Introduction}

Ever since its conception in 1915, Einstein's theory of general relativity (GR) has held up to numerous experimental tests stretching across a wide range of areas. These tests include Solar System observations, such as the perihelion precession of Mercury, as well as binary pulsar observations, such as the orbit period decay of the Hulse-Taylor pulsar~\cite{will-living,TEGP}. In the next couple of decades, these tests and observations will be extended by data from next-generation gravitational wave (GW) detectors~\cite{gwd1,gwd2,gwd3,gwd4,gwd5,gwd6,gwd7,gwd8,gwd9,gwd10,gwd11,gwd12,gwd13,gwd14,gwd15,gwd16,gwd17,gwd18,kent-BDMG-LISA,kent-BDMG-DECIGO,kent-CSGW} (for a recent review of GR tests with ground-based detectors, see~\cite{Yunes:2013dva}). These observations will extend into the \emph{strong-field} regime of gravity, where the gravitational field is nonlinear and dynamical, precisely where tests are currently lacking.

Strong-field tests of gravity have implications to a large range of areas in physics and astrophysics. For example, modified gravity theories that attempt to quantize gravity usually break gravitational parity invariance~\cite{st1,st2,lqg1,lqg2}. Gravitational parity breaking modifies the geometry of spinning black holes (BHs)~\cite{yunespretorius,kent-CSBH} and the propagation of GWs in these backgrounds~\cite{gwd5,gwd14,gwd7,quadratic,kent-CSGW}. Constraining such a departure from the Kerr geometry of GR in the strong-field regime will place constraints on the coupling constants of such theories. These constraints are expected to be orders of magnitude stronger than those one can achieve with Solar System observations.

A recently studied class of modified theories is quadratic gravity (QG)~\cite{nonspinning}. This class departs from GR by adding a dynamical field coupled to all possible curvature squared terms to the Einstein-Hilbert action. The motivation for these modifications comes from similar terms appearing in string theory in a low-energy expansion after compactifying to four dimensions~\cite{st1,st2}, effective field theories of inflation~\cite{eft}, and loop quantum gravity coupled to fermions~\cite{lqg1,lqg2}. Since QG derives as a low-energy expansion of more fundamental theories, it should be viewed as an \emph{effective field theory}, valid up to a cutoff energy scale above which cubic and higher-order curvature invariants cannot be neglected~\cite{kent-CSBH,Dyda:2012rj}. If one does not treat the theory as effective and exceeds the cutoff, ghosts and other instabilities will be nonlinearly generated~\cite{Motohashi:2011pw}, rendering the theory ill-posed.

Two of the more popular QG theories are dynamical Chern-Simons (CS) gravity~\cite{CSreview} and Einstein-Dilaton-Gauss-Bonnet (EDGB) gravity. In the former, the field couples to the Pontryagin invariant, while in the latter it couples to the Gauss-Bonnet invariant. Because of this, dynamical CS gravity is a parity-violating theory, and thus, nonspinning BHs are not modified because they are parity even. Spinning BHs, of course, are not parity even and do acquire corrections~\cite{yunespretorius,konnoBH,kent-CSBH}. BHs in generic QG are different from those in GR already at the non-spinning level~\cite{nonspinning}, and of course also at the spinning level~\cite{Pani}.

For these QG theories to be physically appealing, the BH solutions described above must be stable to perturbations. An instability would imply that BHs generated in gravitational collapse would actually not be realized in nature, if the instability timescale is short enough. QG is obviously linearly stable on a flat background, as higher-order derivatives only arise due to the excitation of the scalar field, which is in turn sourced by the Riemann tensor. A more meaningful test is the study of linear perturbations about solutions to QG that have nontrivial curvature. Such a test was performed on dynamical CS gravity on a Schwarzschild background in~\cite{cardoso-gualtieri,molina,CS,motohashi}. Similar studies were carried out in the context of gravitational radiation in~\cite{pani-DCS-EMRI,quadratic}, but a systematic study of all perturbation modes on generic backgrounds was lacking until now.

In this paper, following the classic work of Isaacson's~\cite{isaacson}, we perform a linear stability analysis on QG in the high-frequency, geometric optics approximation. We concentrate on two particular subclasses of this theory: the even-parity sector, exemplified by EDGB gravity, and the odd-parity sector, exemplified by dynamical CS gravity. We use the high-frequency, geometric optics approximation because to study how perturbative modes or waves propagate on a given background spacetime, the wavelength of the modes has to be much shorter than the curvature length of the background. If this is not the case, the separation of background and wave is ill-defined~\cite{isaacson}. This approximation is sometimes called WKB and is in fact used in many fields, including electromagnetism, quantum mechanics, plasma physics and hydrodynamics.

We here derive general dispersion relations valid for an arbitrary background in modified quadratic gravity, focusing on the even- and odd-parity sectors. We evaluate these relations on nonspinning~\cite{nonspinning} and spinning~\cite{Pani} BH backgrounds that are solutions in these theories. In the odd-parity case, we extend the results of~\cite{CS} by considering spinning BH backgrounds. We consider two particular perturbative regimes: a scalar-dominated and a GW-dominated regime, depending on whether the amplitude of the scalar perturbation is much larger or smaller than the amplitude of the GW perturbation. Our results show that these QG theories are linearly stable in both regimes and for both spherically symmetric and axially symmetric backgrounds in the far field (at distances much farther than the GW wavelength). The speed of coupled gravitational/scalar modes, in the regime where the scalar field perturbation dominates the metric perturbation, is different from that of light, subluminal and superluminal modes, just as in dynamical CS gravity. We argue that this is a generic feature of these theories.

The remainder of this paper presents the mathematical details that back up these results. 
Section~\ref{sec:quad-grav} gives a brief summary of QG, including the modified field equations. 
Section~\ref{sec:lin-stab} outlines the linear stability analysis that is performed and computes the perturbed field equations. 
Section~\ref{sec:disp-rels} finds the dispersion relations needed to analyze the stability of waves within QG and outlines how we study these dispersion relations for a set of examples. 
Section~\ref{sec:egs} studies the solution of the dispersion relations for nonspinning~\cite{nonspinning} and spinning~\cite{Pani} BH spacetimes. 
Section~\ref{sec:concs} concludes by summarizing the results, discussing the implications of said results, and pointing to future possible research.

We will here use the following conventions: we use the metric signature $(-,+,+,+)$; latin letters in index lists stand for spacetime indices; parenthesis and brackets in index lists stand for symmetrization and antisymmetrization, respectively, i.e. $A_{(ab)}=(A_{ab}+A_{ba})/2$ and $A_{[ab]}=(A_{ab}-A_{ba})/2$; we use geometric units with $G=c=1$.

\section{Quadratic Gravity}
\label{sec:quad-grav}

QG can be described by an action containing all possible quadratic, algebraic curvature scalars with running (i.e. nonconstant) couplings~\cite{nonspinning}
\begin{align}
S&\equiv\int d^4x\sqrt{-g}\left\{\kappa R+\alpha_1f_1(\vartheta)R^2+\alpha_2f_2(\vartheta)R_{ab}R^{ab}\right.
\nonumber \\
&+\alpha_3f_3(\vartheta)R_{abcd}R^{abcd}+\alpha_4f_4(\vartheta)R_{abcd}\text{}^{*}R^{abcd}
\nonumber \\
&\left.-\frac{\beta}{2}\left[\nabla_a\vartheta\nabla^a\vartheta+2V(\vartheta)\right]+\mathcal{L}\mat\right\}.
\end{align}
Here, $g$ stands for the determinant of the metric $g_{ab}$. $R$, $R_{ab}$, $R_{abcd}$, and $\text{}^{*}R_{abcd}$ are the Ricci scalar, Ricci tensor, and the Riemann tensor and its dual, respectively, with the latter defined as 
\be
\text{}^{*}R^a_{~bcd}= \frac{1}{2} \varepsilon_{cd}^{~~ef}R^a_{~bef}\,,
\ee
and $\varepsilon^{abcd}$ the Levi-Civita tensor. The quantity $\mathcal{L}_{\mat}$ is the external matter Lagrangian, $\vartheta$ is a field, $f_{i}(\vartheta)$ are functionals of this field, $(\alpha_i,\beta)$ are coupling constants, and $\kappa=1/(16\pi)$. We assume that all quadratic terms are coupled to the same field. All other quadratic curvature terms are linearly dependent, such as the Weyl tensor squared. Terms proportional to derivatives of the curvature can be integrated by parts to obtain the action shown above.

QG contains some well-studied specific theories. For example, $(\alpha_1,\alpha_2,\alpha_3,\alpha_4) = (\alpha_\EDGB, -4\alpha_\EDGB, \alpha_\EDGB,0)$ and $(f_1,f_2,f_3,f_4)=(e^{\vartheta},e^{\vartheta},e^{\vartheta},0)$ correspond to EDGB gravity, where $\alpha_{\EDGB}$ is the EDGB coupling constant and $\vartheta$ is the dilaton. Another example is $(\alpha_1,\alpha_2,\alpha_3,\alpha_4) = (0, 0, 0,\alpha_\CS/4)$ and $(f_1,f_2,f_3,f_4)=(0,0,0,\vartheta)$, which corresponds to dynamical CS gravity, where $\alpha_{\CS}$ is the CS coupling parameter and $\vartheta$ is the CS (axionlike) field. EDGB gravity is constrained most strongly by low-mass x-ray binary observations, $\sqrt{|\alpha_\EDGB|} < 1.9 \times 10^5$cm~\cite{kent-LMXB}, which is 6 orders of magnitude stronger than Solar System bounds~\cite{amendola}. The proposed bound on EDGB with future GW observations is discussed in~\cite{quadratic,kent-LMXB}, where the authors show that space-borne GW interferometers, such as eLISA~\cite{elisa} and DECIGO~\cite{decigo}, should be able to place stronger constraints than the bound mentioned above. On the other hand, dynamical CS theory is most strongly constrained from Solar System~\cite{alihaimoud} and table-top experiments~\cite{kent-CSBH}, $\sqrt{|\alpha_\CS|} < 10^{13}$ cm, while again future GW observations will allow much stronger constraints~\cite{kent-CSGW}.  

In dynamical QG $f_i(\vartheta)$ is some function of the dynamical scalar field $\vartheta$, with potential $V(\vartheta)$. We assume $\vartheta$ is at the minimum of the potential and thus Taylor expand $f_i(\vartheta)=f_i(0)+f'_i(0)\vartheta+\mathcal{O}(\vartheta^2)$ about small perturbations from the minimum (assumed here to be at zero), where $f_i(0)$ and $f'_i(0)$ are constants. The $\vartheta$-independent terms, proportional to $f_i(0)$, lead to a theory with a minimally coupled field, where the latter does not interact with the curvature invariants. In the dynamical CS and the EDGB cases, these invariants are topological, and the $f_{i}(0)$ terms do not modify the field equations. Since we will here concentrate on these theories, the $f_{i}(0)$ are irrelevant and will be neglected. Instead, we concentrate on on the $f'_{i}(0)$ terms, which can be modeled by letting $f_i(\vartheta)=c_i\vartheta$. Reabsorbing the constants $c_i=f'_i(0)$ into $\alpha_i$, such that $\alpha_if_i(\vartheta)\rightarrow\alpha_i\vartheta$, the field equations are then~\cite{nonspinning,quadratic}
\begin {align}
G_{ab}&+\frac{\alpha_1}{\kappa}\mathcal{H}_{ab}^{(\vartheta)}+\frac{\alpha_2}{\kappa}\mathcal{I}_{ab}^{(\vartheta)}
\nonumber \\
&+\frac{\alpha_3}{\kappa}\mathcal{J}_{ab}^{(\vartheta)}+\frac{\alpha_4}{\kappa}\mathcal{K}_{ab}^{(\vartheta)}=\frac{1}{2\kappa}\left(T_{ab}^\mathrm{mat}+T_{ab}^{(\vartheta)}\right), \label{FE}
\end{align}
where 
\begin{equation}
T_{ab}^{(\vartheta)}=\beta\left[\nabla_a\vartheta\nabla_b\vartheta-\frac{1}{2}g_{ab}\left(\nabla_c\vartheta\nabla^c\vartheta-2V(\vartheta)\right)\right]
\end{equation}
is the scalar field stress-energy tensor and
\begin{align}
\mathcal{H}_{ab}^{(\vartheta)}&\equiv -4\nabla_{(a}\vartheta\nabla_{b)}R-2R\nabla_{(a}\nabla_{b)}\vartheta
\nonumber \\
&+g_{ab}\left(2R\nabla^c\nabla_c\vartheta+4\nabla^c\vartheta\nabla_cR\right)
\nonumber \\
&+\vartheta\left[2R_{ab}R-2\nabla_a\nabla_bR-\frac{1}{2}g_{ab}\left(R^2-4\square R\right)\right],
\end{align}
\begin{align}
\mathcal{I}_{ab}^{(\vartheta)}&\equiv -\nabla_{(a}\vartheta\nabla_{b)}R-2\nabla^c\vartheta\left(\nabla_{(a}R_{b)c}-\nabla_cR_{ab}\right)
\nonumber \\
&+R_{ab}\nabla_c\nabla^c\vartheta-2R_{c(a}\nabla^c\nabla_{b)}\vartheta
\nonumber \\
&+g_{ab}\left(\nabla^c\vartheta\nabla_cR+R^{cd}\nabla_c\nabla_d\vartheta\right)
\nonumber \\
&+\vartheta\left[2R^{cd}R_{acbd}-\nabla_a\nabla_bR+\square R_{ab}\right.
\nonumber \\
&\left.+\frac{1}{2}g_{ab}\left(\square R-R_{cd}R^{cd}\right)\right],
\\
\mathcal{J}_{ab}^{(\vartheta)}&\equiv -8\nabla^c\vartheta\left(\nabla_{(a}R_{b)c}-\nabla_cR_{ab}\right)+4R_{acbd}\nabla^c\nabla^d\vartheta
\nonumber \\
&-\vartheta\left[2\left(R_{ab}R-4R^{cd}R_{acbd}+\nabla_a\nabla_bR-2\square R_{ab}\right)\right.
\nonumber \\
&\left.-\frac{1}{2}g_{ab}\left(R^2-4R_{cd}R^{cd}\right)\right],
\\
\mathcal{K}_{ab}^{(\vartheta)}&\equiv 4\nabla^c\vartheta\varepsilon_{c~e(a}^{~d}\nabla^eR_{b)d}+4\nabla_d\nabla_c\vartheta\text{}^{*}R_{(a~b)}^{~~c~d}.
\end{align}
Variation of the action with respect to $\vartheta$ yields the scalar field equations
\begin{align}
\beta\square\vartheta-\beta\frac{dV}{d\vartheta}&=-\alpha_1R^2-\alpha_2R_{ab}R^{ab}
\nonumber \\
&-\alpha_3R_{abcd}R^{abcd}-\alpha_4R_{abcd}\text{}^{*}R^{abcd}.\label{SFE}
\end{align}

\section{Linear Stability Analysis}
\label{sec:lin-stab}

We study the modified field equations in perturbation theory, decomposing the full, spacetime metric into 
\begin{equation}
g_{ab}=\bar{g}_{ab}+\xi' g^{M}_{ab}+ \epsilon h_{ab}^\GR + \epsilon \xi' h_{ab}^\QGpert + \mathcal{O}(\epsilon^2, \xi'^2) 
\label{eq-gab-decomp}
\end{equation}
and the full scalar field into
\begin{equation}
\vartheta=  \xi'^{1/2} \bar{\vartheta}+\epsilon \xi'^{1/2} \delta \vartheta + \mathcal{O}(\epsilon^2,\xi').
\label{eq-theta-decomp}
\end{equation}
$\bar{g}_{ab}$ is a stationary GR solution, $g^{M}_{ab}$ is a stationary, QG modification to this solution, and $h_{ab}^\GR$ and $h_{ab}^\QGpert$ are GR and QG perturbations away from these (background) solutions $\bar g_{ab}+\xi' g^M_{ab}$. $\bar{\vartheta}$ is a stationary solution to the unperturbed field equations and $\delta \vartheta$ is a small perturbation away from this background  field. The book-keeping parameters $\epsilon$ and $\xi'$ denote the order of the perturbation and QG effect respectively. In particular, the latter denotes the order in the coupling parameter $\xi\equiv\alpha^2_i/(\beta\kappa)$, which appears frequently in QG theories. 

All quantities computed in this paper will depend on $g_{ab}$ and $\vartheta$, such as the Riemann tensor, and thus, they can be also decomposed into
\begin{equation}
A=\sum_{n,m} \epsilon^n\xi'^{m/2} A^{(n,m)},
\end{equation}
where $A^{(n,m)}$ is assumed independent of $\epsilon$ and $\xi'$. 
With this decomposition, $\bar{g}_{ab}$, $g^{M}_{ab}$, $h_{ab}^\GR$, $h_{ab}^\QGpert$, $\bar{\vartheta}$ and $\delta \vartheta$ can be written as $g^{(0,0)}$, $g^{(0,2)}$, $g^{(1,0)}$, $g^{(1,2)}$, $\vartheta^{(0,1)}$ and $\vartheta^{(1,1)}$, respectively. For conventional reasons and ease of reading, however, we will continue to decompose the metric and the scalar field as in Eqs.~\eqref{eq-gab-decomp} and \eqref{eq-theta-decomp}. Henceforth, we work to leading order in $\epsilon$ and $\xi'$.

We consider a background metric $\bar g_{ab}+\xi' g^M_{ab}$ that is a vacuum solution to the modified field equations to leading order in $\xi'$. A trivial example of such a solution is the Minkowski spacetime (in this case, the $\xi'$ perturbation vanishes), while more complicated ones are the BH spacetimes found in~\cite{nonspinning} and~\cite{Pani}. In this paper, we will work with a generic background metric $\bar g_{ab}+\xi' g^M_{ab}$ and only later specialize to these BH solutions. The solutions have been shown to actually represent slowly-rotating BHs and not naked singularities in~\cite{nonspinning,Hansen:2013owa}.  

We seek plane-wave solutions to the perturbed metric and scalar field of the form
\begin{align}
h_{ab}=A_{ab}(t,x^{j})e^{i\varphi (t,x^{k})/\epsilon_{\varphi}}\,,
\\
\delta \vartheta = B(t,x^{j})e^{i\varphi (t,x^{k})/\epsilon_{\varphi}}\,.
\end{align}
We further impose the geometric optics approximation, where we require that the phase $\varphi$ varies much faster than the amplitudes $(A_{ab},B)$. This is enforced by requiring that the geometric optics, order-counting parameter $\epsilon_{\varphi}$ be much smaller than unity. With the above ansatz, we are also restricting this analysis to steady-state solutions, hence the scalar field and metric perturbation have the same phase.

The modified field equations and equation of motion for the scalar field can now be expanded trivariately in $\epsilon<<1$, $\xi'<<1$, and $\epsilon_{\varphi}<<1$. The dominant term in the expansion will be $\mathcal{O}(\epsilon,\xi',\epsilon^{-2}_{\varphi})$. We will discard all higher-order terms, and thus, our analysis will not be valid in the non-linear regime $\xi' \sim 1$ by construction. 

\subsection{Perturbed Scalar Field Equations}

The only nonvanishing contribution to the expansion of the left-hand side of the scalar field evolution equation [Eq.~\eqref{SFE}] to $\mathcal{O}(\epsilon,\xi')$ is
\begin{align}
(\square\vartheta)^{(1,1)} &=\bar\square\delta\vartheta - h^{ab}_\GR \bar\nabla_{a}\bar\nabla_{b}\bar\vartheta \nonumber \\
& - \frac{1}{2}\bar g^{ab}[\bar g^{cd}(\partial_{a}h_{bd}^\GR+\partial_{b}h_{da}^\GR-\partial_{d}h_{ab}^\GR)
\nonumber \\
&-h^{cd}_\GR (\partial_{a}\bar g_{bd}+\partial_{b}\bar g_{da}-\partial_{d}\bar g_{ab})]\partial_{c}\bar\vartheta \nn \\
&=\bar\square\delta\vartheta + \mathcal{O}(\epsilon_\varphi^{-1}).
\label{expeps}
\end{align}
$\bar\nabla_a$ and $\bar\square$ denote the covariant derivative and D'Almbertian operator associated with $\bar g_{ab}$, respectively. We are only keeping leading-order terms in $\epsilon_{\varphi}$ and so only terms proportional to second derivatives of a perturbation are kept in the last equality. 

The right-hand side of Eq.~\eqref{SFE} depends on the Riemann tensor which, when expanded about a given background, is given by~\cite{Wald} 
\begin{equation}
R_{abc}{}^d=\bar{R}_{abc}{}^d -2\bar{\nabla}_{[a|}C^d{}_{b]c}+2C^e{}_{c[a|}C^d{}_{b]e}.\label{exp}
\end{equation}
$\bar R_{abc}{}^d$ is the Riemann tensor associated with $\bar g_{ab}$. A similar definition applies to $\bar R_{ab}$ and $\bar R$. The tensor field $C^{c}_{~ab}$ is defined as
\begin{equation}
C^{c}_{~ab}=\frac{1}{2}g^{cd}(\bar{\nabla}_{a}g_{bd}+\bar{\nabla}_{b}g_{ad}-\bar{\nabla}_{d}g_{ab}).
\end{equation}

Using Eq.~\eqref{exp}, the only non-vanishing contribution to the expansion of the right-hand side of Eq.~\eqref{SFE} to $\mathcal{O}(\epsilon,\xi'^0)$ depends on 
\begin{align}
(R_{abcd}R^{abcd})^{(1,0)} &=4\bar{R}^{abcd} \bar\nabla_{a}\bar\nabla_{c}h_{bd}^\GR  + \mathcal{O}(\epsilon_\varphi^{-1}),\label{R^2}
\\ 
(R_{abcd}\text{}^{*} R^{abcd})^{(1,0)} &=2\bar{R}_{abcd}\bar\epsilon^{cdef} \bar\nabla^{a}\bar\nabla_{e}h^{b}_{f, \GR}  + \mathcal{O}(\epsilon_\varphi^{-1}),\label{pont}
\\
(R_{ab}R^{ab})^{(1,0)} &=0,\label{tensor2}
\\
(R^{2})^{(1,0)} &=0\,.\label{scalar2}
\end{align}
We are only considering vacuum solutions, so the background Ricci tensor $\bar R_{ab}$ and scalar $\bar R$ both vanish. Thus, there is no perturbation to the Ricci tensor and scalar squared to the order considered. 

Using Eqs.~\eqref{expeps}, \eqref{R^2}, \eqref{pont}, \eqref{tensor2}, and \eqref{scalar2} in Eq.~\eqref{SFE}, we find the perturbed scalar field equations to leading order in the geometric optics approximation
\begin{equation}
\beta \bar\square\delta\vartheta =-2\bar R_{abcd}\left( 2\alpha_{3}\bar\nabla^{a}\bar\nabla^{c}h^{bd}_\GR+\alpha_{4}\bar\epsilon^{cdef}\bar\nabla^{a}\bar\nabla_{e}h^{b}_{f, \GR}\right).
\end{equation}

Let us now separate the even and odd sectors of QG theories. That is, we now specialize to the even-parity subclass of theories with $(\alpha_{3},\alpha_{4})=(\alpha_{3},0)$ and the odd-parity subclass of theories with $(\alpha_{3},\alpha_{4})=(0,\alpha_{4})$. Defining the four-dimensional wave vector $k_{a}=(\partial_{a}\varphi)/\epsilon_{\varphi}$, and noting that $\partial_d h_{ab}^\GR=h_{ab}^\GR k_d$ to leading order in $\epsilon_{\varphi}$, the perturbed scalar field equations are then
\begin{equation}
(\delta\vartheta_{3})k_{a}k^{a}=-4\bar R^{abcd}h_{bd}^\GR k_{a}k_{c}, \label{psfe3}
\end{equation}
\begin{equation}
(\delta\vartheta_{4})k_{a}k^{a}=-2\bar R_{abcd}\bar\epsilon^{cdef}h^{b}_{f, \GR}k^{a}k_{e}, \label{psfe4}
\end{equation}
where we have rescaled the background scalar field and its perturbation via
\begin{align}
\bar\vartheta_{A} &=\frac{\alpha_{A}}{\beta}\bar\vartheta_{A}, \label{expsf1}
\\
\delta\vartheta_{A} &=\frac{\alpha_{A}}{\beta}\delta\vartheta_{A}\,, \label{expsf2}
\end{align}
with $A=3$ or $4$. Equation~\eqref{psfe4} matches the perturbed scalar field equation for dynamical CS modified gravity found in~\cite{CS}, while Eq.~\eqref{psfe3} is new. The main difference between these two equations is in the appearance of the Levi-Civita tensor associated with the background GR metric in the right-hand side (source) of Eq.~\eqref{psfe4}. This shows clearly that dynamical CS gravity excites modifications to the spectrum of perturbations only for parity-odd backgrounds. 

\subsection{Perturbed Gravitational Field Equations}

Let us first analyze the right-hand side of Eq.~\eqref{FE}. Recall we are only considering vacuum solutions and so $T^{\mat}_{ab}$ vanishes. The scalar field stress-energy tensor $T^{(\vartheta)}_{ab}$ only depends on first derivatives of the scalar field. Thus, $T^{(\vartheta)}_{ab}$ goes as $\epsilon_{\varphi}^{-1}$ to lowest order in $\epsilon_{\varphi}$. The left-hand side of Eq.~\eqref{FE} has terms that depend on the second derivatives of the scalar field, which go as $\epsilon_{\varphi}^{-2}$. Only keeping lowest-order terms in $\epsilon_{\varphi}$, the right-hand side of Eq.~\eqref{FE} is then zero.

Before analyzing the left-hand side of Eq.~\eqref{FE}, let us first make some simplifications. As before, the background Ricci scalar and tensor vanish, since we are only considering vacuum solutions. The $\mathcal{O}(1,0)$ Ricci scalar and tensor also vanish, because they are proportional to $\bar\square h^{\GR}_{ab}$, which vanishes because $h^{\GR}_{ab}$ must satisfy the ${\cal{O}}(1,0)$ perturbed (Einstein) field equations without a source. 

With these simplifications, the perturbed field equations are 
\begin{align}
\frac{\kappa}{4}&\left(\bar g^{cd}\tilde R_{acbd}+g^{cd}_{M} \tilde R_{acbd}\right)
\nonumber \\
=& -\alpha_3\left[\tilde R_{acbd}\bar\nabla^c\bar\nabla^d\bar\vartheta+\bar R_{acbd}\bar\nabla^c\bar\nabla^d\left(\delta\vartheta\right)\right]
\nonumber \\
& -\alpha_4\left[\text{}^{*}\tilde R_{(a|c|b)d}\bar\nabla^d\bar\nabla^c\bar\vartheta+\text{}^{*}\bar R_{(a~b)}^{~~c~d}\bar\nabla_d\bar\nabla_c\left(\delta\vartheta\right)\right],\label{FE2}
\end{align}
where $\tilde R_{abcd}=R^{(1,0)}_{abcd}+\xi' R^{(1,2)}_{abcd}$. The perturbed Riemann and its dual, as well as the contraction of the GR background metric with the perturbed Riemann, are
\begin{align}
\tilde R_{acbd}&=\left(\tilde h_{[c|d}k_{|a]}k_b-\tilde h_{[c|b}k_{|a]}k_d\right),
\\
\text{}^{*}\tilde R_{(a|c|b)d}&=\frac{1}{2}\bar\varepsilon_{(a|d}^{~~~~ef}\left(\tilde h_{ce}k_{|b)}k_f-\tilde h_{|b)e}k_ck_f\right),
\\
\bar g^{cd}\tilde R_{acbd}&=\frac{1}{2}\tilde h_{ab}k_ck^c.
\end{align}
As with the perturbed Riemann, $\tilde h_{ab}=h^{(1,0)}_{ab}+\xi' h^{(1,2)}_{ab}$.

Let us now specialize the analysis by again separating the even-parity and odd-parity sectors of QG. We thus use the expansions for the scalar field in Eqs.~\eqref{expsf1} and~\eqref{expsf2} and the following rescaling  the modified and perturbed metric:
\begin{align}
g^{M_{A}}_{ab}&=\xi_{A} g^{M_{A}}_{ab}\,,
\\
\tilde h^A_{ab} &= h^{A(1,0)}_{ab}+\xi'_A h^{A (1,2)}_{ab}\,,
\end{align}
where $A=3$ or $4$. The even- and odd-parity perturbed field equations, those proportional to only $\alpha_3$ and $\alpha_4$, are then 
\begin{align}
\tilde h^{3}_{ab}k_ck^c&=-8\xi_3 \bar R_{acbd}\left(\delta\vartheta_3\right)k^c k^d
\nonumber \\
&-8\xi_3 \tilde R^3_{acbd}\bar\nabla^c\bar\nabla^d\bar\vartheta_3-2\xi_3 g^{cd}_{M_3}\tilde R^3_{acbd}, \label{pfe3}
\\
\tilde h^4_{ab}k_ck^c&=-8\xi_4\text{}^{*}\bar R_{(a|c|b)d}\left(\delta\vartheta_4\right)k^ck^d
\nonumber \\
& -8\xi_4\text{}^{*}\tilde R^4_{(a|c|b)d}\bar\nabla^c\bar\nabla^d\bar\vartheta_4 -2\xi_4 g^{cd}_{M_4}\tilde R^4_{acbd}\,, \label{pfe4}
\end{align}
where $\tilde R^A_{acbd}$ denotes the perturbed Riemann proportional to $\tilde h^A_{cd}$, with $A = 3$ or $4$. 

\section{Dispersion Relations}
\label{sec:disp-rels}

We now combine the perturbed scalar and gravitational field equations to obtain dispersion relations in both the even- (EDGB) and odd-parity (dynamical CS) sectors of QG. To achieve this goal, one needs to first replace $h_{ab}^\GR$ in Eqs.~\eqref{psfe3} and~\eqref{psfe4} with $\tilde{h}_{ab}$. This is justified because the difference between $h_{ab}^\GR$ and $\tilde{h}_{ab}$ is of $\mathcal{O}(\xi')$, and thus, it does not affect the final result to the working order. Using the perturbed field equations, together with Eqs.~\eqref{pfe3} and~\eqref{pfe4}, we find the dispersion relations
\begin{align}
\left(k_ak^a\right)^2 &=32\xi_3\bar R^{abcd}\bar R_{bedf}k_ak_ck^ek^f
\nonumber \\
&+\frac{32\xi_3}{(\delta\vartheta_3)}\bar R^{abcd}\tilde R^3_{bedf}k_ak_c\bar\nabla^e\bar\nabla^f\bar\vartheta_3
\nonumber \\
&+\frac{8\xi_3}{(\delta\vartheta_3)}\bar R^{abcd}\tilde R^3_{bedf}k_ak_cg^{ef}_{M_3},
\end{align}
in the even-parity sector and 
\begin{align}
\left(k_ak^a\right)^2&=16\xi_4 \bar\epsilon^{cdef}\bar R_{a~cd}^{~b}\text{}^{*}\bar R_{(b|g|f)h}k^ak_ek^gk^h
\nonumber \\
&+\frac{16\xi_4}{(\delta\vartheta_4)}\bar\epsilon^{cdef}\bar R_{a~cd}^{~b}\text{}^{*}\tilde R^4_{(b|g|f)h}k^ak_e\bar\nabla^g\bar\nabla^h\bar\vartheta_4
\nonumber \\
&+\frac{4\xi_4}{(\delta\vartheta_4)}\bar\epsilon^{cdef}\bar R_{a~cd}^{~b}\tilde R^4_{bgfh}k^ak_eg^{gh}_{M_4}.
\end{align}
in the odd-parity sector. We immediately notice that these relations are not quite dispersion relations because the right-hand side depend on the perturbations. In a slight abuse of nomenclature, however, we will continue to refer to them as dispersion relations~\cite{garfinkle}. 

We can simplify these relations by rewriting the GR background Riemann tensor in terms of the GR background Weyl tensor $\bar C_{abcd}$, since the GR background considered is vacuum. Let us then define the tensors $\bar W_{ab}$ and $\bar S_{ab}$ in analogy with the electric and magnetic parts of the background Weyl tensor
\begin{align}
\bar W_{ac}&=\bar C_{abcd}k^bk^d,
\nonumber \\
\bar S_{ac}&=\text{}^{*}\bar C_{abcd}k^bk^d=\frac{1}{2}\varepsilon_{cd}^{~~ef}\bar C_{abef}k^bk^d.\label{WS}
\end{align}
Note that a further contraction of $\bar S_{ac}$ or $\bar W_{ac}$ with $k_a$ or $k_c$ vanishes due to symmetry. The dispersion relations in terms of these tensors become
\begin{align}
(k_ak^a)^2&=32\xi_3 (\bar W^{ab})(\bar W_{ab})
\nonumber \\
&+32\xi_3 \frac{(\bar W^{ab})}{(\delta\vartheta_3)}\tilde R^3_{acbd}\bar\nabla^c\bar\nabla^d\bar\vartheta_3
\nonumber \\
&+8\xi_3 \frac{(\bar W^{ab})}{(\delta \vartheta_3)}g^{cd}_{M_3}\tilde R^3_{acbd}, \label{dr3}
\end{align}
in the even-parity sector and
\begin{align}
(k_ak^a)^2&=32\xi_4 (\bar S^{ab})(\bar S_{ab})
\nonumber \\
&+32\xi_4\frac{(\bar S^{ab})}{(\delta\vartheta_4)}\text{}^{*}\tilde R^4_{acbd}\bar\nabla^c\bar\nabla^d\bar\vartheta_4
\nonumber \\
&+8\xi_4\frac{(\bar S^{ab})}{(\delta\vartheta_4)}g^{cd}_{M_4}\tilde R^4_{acbd},\label{dr4}
\end{align}
in the odd-parity sector. Note that when $\xi_3$ and $\xi_4$ vanish, the dispersion relations are both null, returning the GR result as expected.

The dispersion relations in Eqs.~\eqref{dr3} and~\eqref{dr4} depend on the amplitude of the incident scalar and gravitational wave. Therefore, in order to make analytic progress, we will define two different perturbative regimes: one in which the magnitude of the scalar perturbation dominates over the magnitude of the metric perturbation (the scalar regime), and another one in which the opposite is true (the GW regime). To achieve this, let us treat the perturbation to the metric as a GW, and decompose it into $+$ and $\times$ polarizations: 
\begin{equation}
A_{ab}\equiv A_+ e^+_{ab}+A_\times e^\times_{ab}\equiv A_{+,\times}e^{+,\times}_{ab}.
\end{equation}
which clearly only affects the magnitude of the perturbation. In what follows, we define and study each of these perturbative regimes in more detail.

Physically, whether a perturbed black hole will be in scalar or GW regime depends on the particular type of perturbation considered. For example, consider the case where the perturbations are generated from the quasi-circular inspiral of two black holes into each other (that are far away from the background black hole that is being perturbed). In this case,~\cite{Yagi} computed the far-zone perturbations both in the even- and odd-parity sectors of quadratic gravity. They found that $B_{\rm even} \sim (m/D_{L}) v_{12}$, while $B_{\rm odd} \sim (m/D) v_{12}^{4} \chi$, where $m$ is the total mass of the binary, $D_{L}$ is the luminosity distance from the binary's center of mass to the field point (in this case, the location of the background black hole), $v_{12}$ is the magnitude of the orbital velocity of the binary and $\chi$ is the dimensionless magnitude of the spin angular momenta of either black hole binary component. 

To determine which regime dominates, these scalar amplitudes are to be compared with the amplitude of the GR gravitational wave metric perturbation, which is simply $A_{+,\times} \sim  (m/D) v_{12}^{2}$. We then see that, in this example, $B_{\rm even} \gg A_{+,\times}$ (scalar regime) but $B_{\rm odd} \ll A_{+,\times}$ (GW regime), because during the inspiral $v_{12} \ll 1$. Of course, this is just an example, since in general, the magnitude of $A_{+,\times}$ and $B$ will depend on the particular system considered (e.g. during a supernovae, $B$ and $A_{+,\times}$ will have a different scaling). Thus, to determine whether one is in one regime or the other, the particular scenario that is producing the perturbation must be carefully analyzed. 

\subsection{Scalar Regime}

Let us define the scalar regime as that in which the amplitude of the scalar field perturbation is much larger than the amplitude of the metric perturbation, i.e.
\begin{equation}
B \gg A_{+,\times}\,.
\end{equation}
\\ \indent
In this limit the second and third terms in each dispersion relation is dominated by the first term giving the following dispersion relation:
\begin{equation}
(k_ak^a)^2=32\xi_{A} (\bar S^{ab})(\bar S_{ab})\,, \label{sr}
\end{equation}
where $A = 3$ or $4$. This relation has been obtained using the identity $S_{ab} S^{ab} = W_{ab} W^{ab}$, which is only valid to leading order in $\xi'$ (see the Appendix for a full discussion). Notice that this relation is the same for both parity sectors (with the obvious change of coupling constant). Notice also that, in the scalar regime, the perturbations disappear from the right-hand side, and thus, Eq.~\eqref{sr} is actually a true dispersion relation. 

In the dynamical CS case, the general dispersion relation of Eq.~\eqref{dr4} reduces to that of Eq.~\eqref{sr} with $A=4$ when the GR background is spherically symmetric. This is because in that case the background scalar field vanishes $\bar{\vartheta}_{4} = 0$, since the source to the scalar evolution equation also vanishes. This then implies that the CS correction to the GR background metric also vanishes $g_{cd}^{M_{4}} = 0$, since there is no scalar field to source a modification. One then sees that the second and third lines of Eq.~\eqref{dr4} vanish identically, reducing to Eq.~\eqref{dr4}. This agrees with the findings of~\cite{CS}.

Already at this stage, we can make some interesting observations about the modified dispersion relations in the scalar regime. For any Petrov type D spacetime~\cite{Stephani:2003tm}, 
\be
\bar{S}_{ac} = {}^{*}\bar{C}_{abcd} k^{b} k^{d} = \lambda k_{a} k_{c}\,,
\ee
for some constant $\lambda$, assuming $k^{a}$ is a principal null direction of the Weyl tensor. It follows then that $\bar{S}_{ab} \bar{S}^{ab} = \lambda^{2} (k_{a} k^{a})^{2} = {\cal{O}}(\xi'^{2})$ for such spacetimes, because $k_{a} k^{a} = {\cal{O}}(\xi')$, since $h_{ab}^{\GR}$ satisfies the linearized Einstein equations. Thus, the modified dispersion relation in Eq.~\eqref{sr} reduces to the GR expression. In particular, background GR spacetimes that represent BHs, such as the Schwarzschild and Kerr metrics, are Petrov type D with principal null directions in the radial direction. Therefore, it follows that for radial modes propagating in such BH backgrounds, QG leads to the same dispersion relation as GR~\cite{garfinkle}.   

A trivial generalization of the above result is that the modified dispersion relations reduce to the GR one for any Petrov type N spacetime with waves propagating along the null directions of the Weyl tensor. This follows from the fact that for such spacetimes, $\bar{C}_{abcd} k^{c} = 0$~\cite{Stephani:2003tm}.

\subsection{GW Regime}

Let us define the GW regime as that in which the amplitude of the metric perturbation is much larger than the amplitude of the scalar field perturbation, i.e.
\begin{equation}
B \ll A_{+,\times}
\end{equation}
In this limit the second and third terms in the dispersion relations dominate, and one finds 
\begin{align}
(k_ak^a)^2&=16\xi_3 \frac{A^3_{+,\times}}{B_3} \bar W^{ab}e^{+,\times}_{ab}k_{c}k_{d}\bar\nabla^c\bar\nabla^d\bar\vartheta_3
\nonumber \\
&+4\xi_3 \frac{A^3_{+,\times}}{B_3}\bar W^{ab}e^{+,\times}_{ab}k_{c}k_{d}g_{M_3}^{cd}, \label{gwr3}
\end{align}
in the even-parity sector and
\begin{align}
(k_ak^a)^2&=16\xi_4 \frac{A^4_{+,\times}}{B_4} \bar S^{ab}\bar\varepsilon_{ad}^{~~~ef}e^{+,\times}_{be}k_ck_f\bar\nabla^c\bar\nabla^d\bar\vartheta_4
\nonumber \\
&+4\xi_4\frac{A^4_{+,\times}}{B_4} \bar S^{ab}e^{+,\times}_{ab}k_{c}k_{d}g_{M_4}^{cd} \label{gwr4},
\end{align}
in the odd-parity sector. The $3$ and $4$ indices on $A_{+,\times}$ and $B$ denote that they are the amplitude or wave tensor associated with $h_3$ and $\vartheta_3$, or $h_4$ and $\vartheta_4$, respectively.

\section{Propagation Speed in BH Backgrounds}
\label{sec:egs}

The dispersion relations we found in the previous section can be solved to find the speed of propagating modes in a specific background that is a solution to QG. To do so we parametrize the four wave vector as 
\begin{equation}
k^a\equiv [\Omega,k_1,k_2,k_3]
\end{equation}
and solve Eq.~\eqref{sr} or Eqs.~\eqref{gwr3} and~\eqref{gwr4} for $\Omega$.

\subsection{Scalar Regime}
The dispersion relations in the scalar regime only depend on the GR background so the Schwarzschild and Kerr metrics can be used for nonspinning and spinning BHs, respectively. Moreover, recall that the dispersion relations are the same for the even- (EDGB) and odd-parity (dynamical CS) sectors of QG, with the obvious change of coupling constant. 
\subsubsection{Non-spinning BH background}
Let us first concentrate on a Schwarzschild background in Schwarzschild coordinates $(t,r,\theta,\phi)$. Keeping only the lowest-order in $\xi'$ correction, one finds
\begin{equation}
\Omega\QG=\Omega\Schw\left[1\pm \frac{12m^3}{r^3}\zeta_{A}^{1/2}\left(1-\frac{k_{1}^{2}}{\Omega\Schw^{2}f^2}\right)\right].\label{OmegaQG}
\end{equation}
Here $\zeta_A=\xi_A/m^4$, $m$ is the physical BH mass defined in terms of the ``bare'' mass $M_0$ as $m=M_0\left[1+\left(49/80\right)\zeta_A\right]$, $f=1-2m/r$, and the GR dispersion result is
\begin{equation}
\Omega\Schw=\pm \frac{1}{f}\sqrt{k_{1}^{2}+fr^2k_{2}^{2}+fr^2 \sin^2(\theta) k_{3}^{2}}.
\end{equation}
Notice that there are four independent solutions because the dispersion relations in Eqs.~\eqref{dr3} and~\eqref{dr4} are quartic polynomials in $\Omega$. This matches the result found in~\cite{CS} for dynamical CS gravity, as expected. Interestingly, we find the same exact GW speed for the EDGB case, modulo the coupling constant. This stems from the identity in the Appendix.

Note when dealing with an ingoing or outgoing spherical wavefront ($k_{2,3}=0$), there is no QG correction. This is because the QG modification to the propagation speed is proportional to the difference $(\Omega_{\Schw}^{2} - k_{1}^{2})$, which identically vanishes when $k_{2,3} = 0$. Therefore, in the far-zone limit (a distance much farther than the GW wavelength), where any GW observations would be made and all waves approach a spherical wavefront, the four distinct solution of Eq.~\eqref{OmegaQG} degenerate into the GR prediction exactly. Thus, distinguishing a GR BH from a QG BH based on wave speed would be impossible.
\subsubsection{Spinning BH}
Let us now concentrate on a Kerr background, expanded to $\mathcal{O}(\chi^2)$ with $\chi \equiv a/m$, where $a$ is the Kerr spin parameter and $m$ is the physical BH mass. Keeping only the lowest order in $\xi'$ correction, one finds 
\begin{equation}
\Omega =\Omega\QG + \chi \; \Omega\QGa + \chi^2 \; \Omega\QGaa,
\label{spin-sol}
\end{equation}
where $\Omega\QG$ is the solution in the nonspinning case, already given in Eq.~\eqref{OmegaQG}. $\Omega\QGa$ and $\Omega\QGaa$ are $\mathcal{O}(\chi)$ and $\mathcal{O}(\chi^2)$ corrections, respectively, which are given by
\begin{align}
\Omega\QGa &=\Omega\Kerra \left[1-\frac{12m^2}{r^2}\zeta_{A}^{1/2}\right],
\\
\Omega\QGaa &= \Omega\Kerraa + \zeta_{A}^{1/2}\Omega\Deltaaa,
\end{align}
where 
\begin{align}
\Omega\Kerra &=\pm\frac{2m^2k_3\sin^2(\theta)}{rf},
\\
\Omega\Kerraa &=\pm\frac{m^2}{2r^2f^2\left(\Omega\Schw f\right)} \left[\Omega\Schw^2 f^2 \cos^2(\theta)\left(1-\frac{4m}{r}\right) \right. \nn \\
& \left.-k_1^2+k_3^2 r^2 f \sin^4(\theta) \right],
\\
\Omega\Deltaaa&=\pm\frac{6m^5}{r^5f^2\left(\Omega\Schw f\right)}\left[\left(1-\frac{k_1^2}{\Omega\Schw^2 f^2}\right)\right.
\nonumber \\
&\left.\times\left(k_1^2-k_3^2r^2f\sin^4(\theta)-2D_{\mp}\Omega\Schw^2f^2\cos^2(\theta)\right)\right.
\nonumber \\
&\left.+E\left(2\Omega\Schw^2f^2\sin^2(\theta)+4k_3^2r^2f\sin^2(\theta)\right)\right],
\end{align}
$D_{-}=\frac{m}{r}$, $D_{+}=(1-\frac{m}{r})$, $E=-1$ for the $D_{-}$ case and $E=+1$ for the $D_{+}$ case. Here $(t,r,\theta,\phi)$ are standard Boyer-Lindquist coordinates.
The result in GR is
\begin{equation}
\Omega\GR =\Omega\Schw + \chi \; \Omega\Kerra + \chi^2 \; \Omega\Kerraa.
\end{equation}

We can once again analyze this solution in the far-field limit, where waves are essentially radial ($k_{2,3}=0$). As in the nonspinning BH background case, the QG correction that is $\chi$ independent identically vanishes, as also does the linear-in-$\chi$ correction. However, at ${\cal{O}}(\chi^{2})$ the QG modification does not vanish. The correction is 
\be
\Omega\Deltaaa (k_{2,3}=0)=\mp 12 \left(\frac{m}{r}\right)^5  k_1 \sin^2(\theta) + {\cal{O}}\left(\frac{m^{6}}{r^{6}}\right),
\ee
and
\be
\Omega\Kerraa (k_{2,3}=0)=\mp \frac{1}{2} \left(\frac{m}{r}\right)^{2} k_1 \sin^2(\theta) + {\cal{O}}\left(\frac{m^{3}}{r^{3}}\right).
\ee
Notice again that the four independent solutions of Eq.~\eqref{spin-sol} degenerate into two independent solutions in the far-field limit. Notice also that the QG correction decays much faster with distance from the BH, and, in turn, will be a much weaker effect in the far zone compared to the GR term. Thus, even in the spinning BH case, the effect produced by QG on the wave speed would be essentially impossible to distinguish from the GR prediction.
\subsection{GW Regime}
For simplicity we only analyze radial waves ($k_{2,3}=0$) in the far-zone regime. In this regime, the polarization tensor is simple~\cite{Thorne}. The background scalar field and QG modification to the metric are given in~\cite{nonspinning} and~\cite{Pani} for the nonspinning and spinning cases, respectively. In this case, however, recall that the dispersion relations are different for the even- and odd-parity sectors of QG. 

\subsubsection{Non-spinning BH}
Let us first consider the odd-parity (dynamical CS) case. For nonspinning backgrounds, the background scalar field is zero, and thus, the background QG modification also vanishes. As before, then, the QG modification to the dispersion relation vanishes in the far-zone and we have $\Omega=\Omega\GR = \pm k_{1}$. 

In the even-parity (EDGB) case, however, the background scalar field does not vanish. However, expanding the solution in the far-zone we still find that it reduces to the GR solution $\Omega = \Omega\GR=\pm k_1$. This is because the source term to the modifications of the dispersion relation scale with a high power of $1/r$, even higher than in the scalar regime. Thus, just as in the latter, radial waves propagating in the far-zone are indistinguishable from the GR prediction. 

\subsubsection{Spinning BH}
For spinning BH backgrounds, both the odd-parity, dynamical CS and the even-parity, EDGB background scalar fields are nonvanishing, thus sourcing a QG correction to the background metric. However, when expanding the solution in the far zone, the speed of the propagating modes reduces exactly to the GR prediction $\Omega = \Omega_{\GR} = \pm k_{1}$, to leading order in $m/r$ and irrespective of whether one consider a plus or a cross polarized 

We can see this phenomenon clearly in the $\mathcal{O}(\chi^2)$ term of the solution to the dispersion relation. Such a correction enters first at $\mathcal{O}(m^2/r^2)$, and thus, to leading order in $m/r$ it must be discarded. Notice that this is true regardless of whether one considers the even- or odd-parity dispersion relations. We thus conclude that the speed of propagating modes in QG is the same when considering spinning and nonspinning BH backgrounds, and in both cases identical to the GR prediction in the far field. 

\section{Conclusion}
\label{sec:concs}

In this paper, we have studied the linear stability to high-frequency perturbations of certain background solutions in the far zone. We focused on perturbations of nonspinning~\cite{nonspinning} and spinning~\cite{Pani} BHs, expanding on previous work by including spinning BH backgrounds. We started by considering generic QG theories, but soon after specialized to even-parity and odd-parity subclasses of QG theories. We derived dispersion relations in these two subclasses and solved them in two special regimes: the scalar-dominated and GW-dominated regimes, defined by whether the amplitude of the scalar to the GW perturbations is much larger or smaller than unity. In all cases, we find that the speed of propagating modes is not equal to that of light, but rather faster or slower, depending on the direction of incidence. These GR corrections, however, are too weak to be measurable by Earth-bound detectors due to its rapid radial fall off.

Our work extends all previous work on metric and scalar perturbations in dynamical CS gravity. In addition to the work in~\cite{CS}, Motohashi and Suyama~\cite{motohashi} have performed a perturbation analysis of this theory assuming a spherically symmetric background spacetime and using a spherical harmonic decomposition. They found that all modes propagate at the speed of light if the background scalar field vanishes. If one takes the high-frequency limit of their analysis, their wave ansatz corresponds to a spherical wavefront, and our result automatically reduces to theirs (and also those of \cite{CS}). It would be interesting to extend the analysis of~\cite{motohashi} to see if any ghost mode exists in QG assuming a spherically symmetric background.

We have omitted a full analysis of the dispersion relations in the GW regime, Eqs.~\eqref{gwr3} and ~\eqref{gwr4}, on the backgrounds that we studied. The dispersion relations are directly dependent on the amplitudes of the metric and scalar field perturbations. Due to this dependence, it is not obvious whether the scalar and GWs will be generically stable. The analysis we carried out in the far field, however, suggests there are no imaginary terms in the solution and so the waves should be stable, but it is possible that taking this limit suppresses imaginary terms.

Our stability analysis was performed only to linear order in $\epsilon$ and $\epsilon_{\varphi}$. Instabilities, if they exist, commonly are found at higher than linear order in the perturbation. Another possible extension would be to perform a stability analysis to second or even higher order in $\epsilon$ and $\epsilon_{\varphi}$. However, going to higher order in $\xi'$ would not be possible, without including terms of higher order in the curvature in the quadratic action.

\acknowledgements

The authors thank Frans Pretorius and David Garfinkle for very useful suggestions and comments. Some calculations used the computer algebra systems MAPLE, in combination with the GRTensor II package~\cite{GRT}. N. Y. acknowledges support from NSF grant PHY-1114374, as well as support provided from the National Aeronautics and Space Administration from grant NNX11AI49G, under subaward 00001944.

\appendix
\section{Weyl-Riemann Identity}

Here we show that
\begin{equation}
\bar W^{ac}\bar W_{ac}=\bar S^{ac}\bar S_{ac} + {\cal{O}}(\xi').
\end{equation}
Using the definition of $\bar W^{ac}$ and $\bar S^{ac}$ in Eq.~\eqref{WS} we have
\begin{align}
\bar W^{ac} \bar W_{ac}&=\bar C_{abcd} \bar C^{aecf} k^b k^d k_e k_f,
\\
\bar S^{ac} \bar S_{ac}&=\frac{1}{4}\epsilon_{cdef}\epsilon^{chij}\bar C^{abef} \bar C_{agij} k_b k^d k^g k_h.
\end{align}
Contracting the Levi-Civita tensors
\begin{align}
4\bar S^{ac} \bar S_{ac} &= -6\delta^{d}_{[h}\delta^{i}_{e}\delta^{j}_{f]}\bar C_{agij} \bar C^{abef} k_b k^d k^g k_h
\nonumber \\
&=2\bar C^{abij}\bar C_{agij}k_bk^gk^hk_h
 \nonumber \\
&+2\bar C^{abjh}\bar C_{agij}k_bk^ik^gk_h
 \nonumber \\
&+2\bar C^{abhi}\bar C_{agij}k_bk^jk^gk_h
\nonumber \\
&=2\bar C^{abij}\bar C_{agij}k_bk^gk^hk_h-4W^{ab}W_{ab}.
\end{align}
Notice that the first term contains the term $k^ak_a$, which vanishes in GR because of the ${\cal{O}}(\epsilon,\xi'^{0})$ perturbed Einstein equations. Thus, $k^{a}k_{a} = 0 + {\cal{O}}(\xi')$. Therefore, the first term can be ignored to leading order in $\xi'$ and we find
\begin{equation}
\bar W^{ac} \bar W_{ac} = \bar S^{ac} \bar S_{ac} + \mathcal{O}(\xi').
\end{equation}

\bibliography{biblio}
\end{document}